\documentclass[showpacs,preprintnumbers,superscriptaddress,prl,twocolumn]{revtex4}
\usepackage{subfigure}
\usepackage{graphicx,dcolumn,bm,color,latexsym,amssymb}
\usepackage{amsmath}
\usepackage{amsfonts}
\usepackage{mathrsfs}
\usepackage{arydshln}
\usepackage[sans]{dsfont}
\usepackage[latin1]{inputenc}
\usepackage{color}

\definecolor{Blue}{rgb}{0.0,0.0,1}
\definecolor{Red}{rgb}{1,0.0,0.0}
\definecolor{Green}{rgb}{0,0.5,0.0}

\begin{document}

\title{Computational speed-up in a single qudit NMR quantum information processor}

\author{I. A. Silva}\thanks{These authors contributed equally to this work.} \affiliation{Instituto de F\'isica de S\~ ao Carlos, Universidade de S\~ ao Paulo, Caixa Postal 369, 13560-970 S\~ ao Carlos, S\~ ao Paulo, Brazil}
\author{B. \c{C}akmak }\thanks{These authors contributed equally to this work.} \affiliation{Faculty of Engineering and Natural Sciences, Sabanci University, Tuzla, Istanbul, 34956, Turkey}
\author{G. Karpat} \thanks{These authors contributed equally to this work.}\affiliation{Faculdade de Ci\^encias, UNESP - Universidade Estadual Paulista, Bauru, S\~ ao Paulo, 17033-360, Brazil}
\author{E. L. G. Vidoto} \affiliation{Instituto de F\'isica de S\~ ao Carlos, Universidade de S\~ ao Paulo, Caixa Postal 369, 13560-970 S\~ ao Carlos, S\~ ao Paulo, Brazil}
\author{D. O. Soares-Pinto} \affiliation{Instituto de F\'isica de S\~ ao Carlos, Universidade de S\~ ao Paulo, Caixa Postal 369, 13560-970 S\~ ao Carlos, S\~ ao Paulo, Brazil}
\author{E. R. deAzevedo} \affiliation{Instituto de F\'isica de S\~ ao Carlos, Universidade de S\~ ao Paulo, Caixa Postal 369, 13560-970 S\~ ao Carlos, S\~ ao Paulo, Brazil}
\author{F. F. Fanchini} \affiliation{Faculdade de Ci\^encias, UNESP - Universidade Estadual Paulista, Bauru, S\~ ao Paulo, 17033-360, Brazil}
\author{Z. Gedik} \affiliation{Faculty of Engineering and Natural Sciences, Sabanci University, Tuzla, Istanbul, 34956, Turkey}
\date{\today}

\begin{abstract}
Quantum algorithms are known for presenting more efficient solutions to certain computational tasks than any corresponding classical algorithm. It has been thought that the origin of the power of quantum computation has its roots in non-classical correlations such as entanglement or quantum discord. However, it has been recently shown that even a single pure qudit is sufficient to design an oracle-based algorithm which solves a black-box problem faster than any classical approach to the same problem. In particular, the algorithm that we consider determines whether eight permutation functions defined on a set of four elements is positive or negative cyclic. While any classical solution to this problem requires two evaluations of the function, quantum mechanics allows us to perform the same task with only a single evaluation. Here, we present the first experimental demonstration of the considered quantum algorithm with a quadrupolar nuclear magnetic resonance setup using a single four-level quantum system, i.e., a ququart.
\end{abstract}


\maketitle

\textit{Introduction}.-- During the last decades there has been an ever increasing interest in the development novel quantum algorithms to provide a speed-up for the solution of computational tasks over classical algorithms \cite{nielsen,cleve}. Study of quantum algorithms has not only served the purpose of solving certain problems faster than any corresponding classical algorithm but has also opened up the discussion of what actually is the resource of quantum computation. Advances in this research line has not been limited to purely theoretical proposals, as many such quantum algorithms have been demonstrated with several different experimental systems \cite{nielsen,ivan,bianucci}.

At the heart of quantum information science lies the concept of quantum entanglement, which has been regarded as the defining property of the quantum information theory for many years \cite{horodecki}. In the words of Schr\"{o}dinger himself, ``entanglement is not \textit{one} but rather \textit{the} characteristic trait of quantum mechanics''. Indeed, this peculiar property has been demonstrated to provide the magic for many quantum protocols such as the Deutsch algorithm \cite{deutsch,colins}, which distinguishes constant functions from balanced ones, and the Shor algorithm \cite{shor} which finds the prime factors of a given integer.

However, the idea that the entanglement is the one and only fundamental resource of quantum computation has started to change in the last decade. Despite the undeniable importance of entanglement to quantum information theory, novel ways of understanding quantum correlations from different perspectives have emerged. It has been shown that the existence of more general quantum correlations quantified by discord-like measures, even in the absence of any entanglement, might be responsible for the improved performance of some quantum protocols \cite{modi}. Another recent candidate for such improvements is the contextual nature of quantum mechanics \cite{howard}. All in all, the origin of the power of quantum algorithms is still not completely clear yet \cite{nest,vedral,brodutch}.

\begin{figure*}[t]
\includegraphics[scale=0.3]{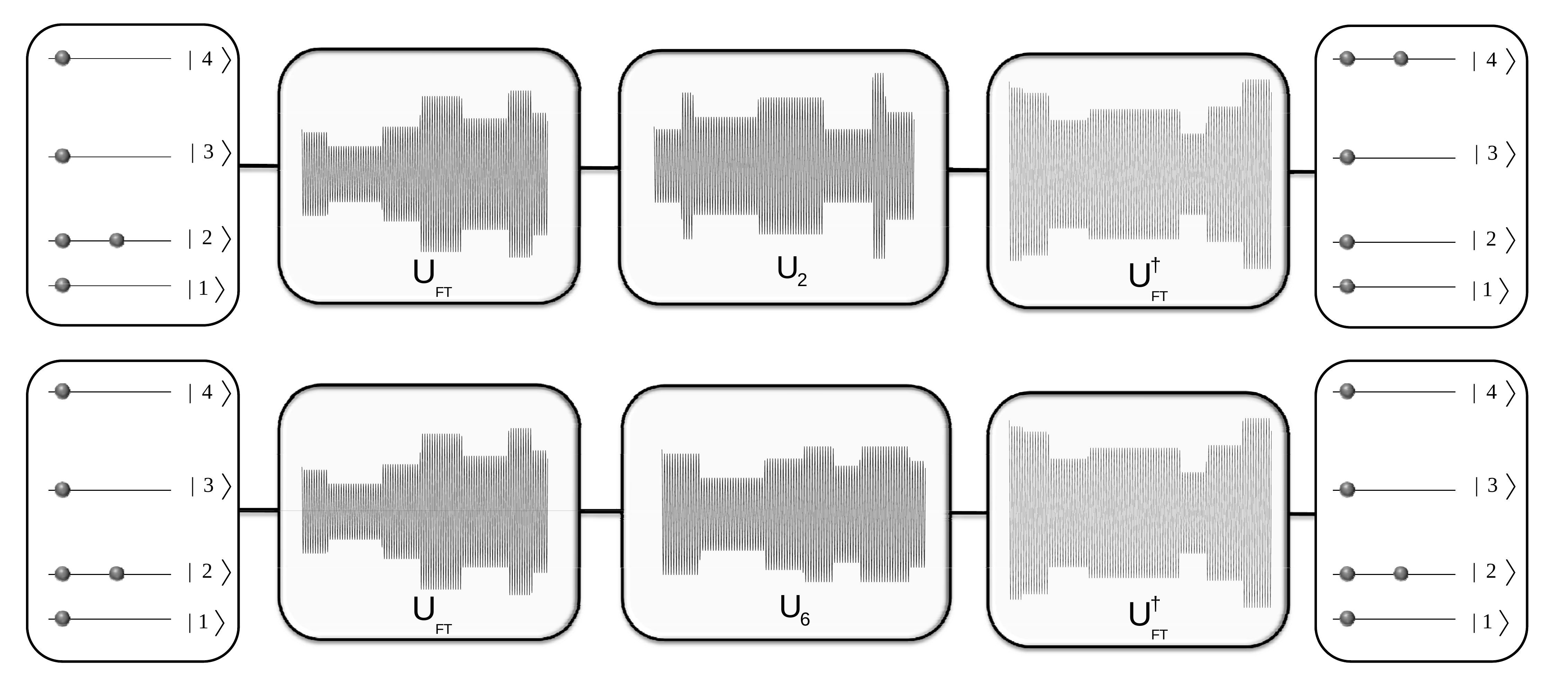} \caption{\label{Circuit} Pictorical view of the experimental implementation: the first boxes represent the level populations of the initial states, the consecutive three boxes represent the optimized rf pulse sequence that implement the quantum gates, and the last boxes represent level populations of the final state. In both cases, the initial state is $|2\rangle$ and, depending on the applied unitary transformation $U_i\;(i = 2, 6)$, the final state can be either $|2\rangle$ or $|4\rangle$. The illustrated rf pulse sequences that implement each operation were optimized using a home-built program based on the strong modulating pulses procedure \cite{fortunato}.}
\end{figure*}

In this work, we consider a novel oracle-based quantum algorithm which, based on a surprisingly simple idea, solves a black-box problem using only a single qudit without any correlation of quantum or classical nature \cite{gedik}. The algorithm deals with the problem of deciding whether chosen $2d$ permutation functions of $d$ objects is positive or negative cyclic with a single query to the black-box rather than two queries required by any corresponding classical algorithm for the solution of the same problem. Here, we present the first experimental demonstration of this algorithm using a room temperature nuclear magnetic resonance (NMR) quadrupolar system.

\begin{figure*}[t]
\includegraphics[scale=0.7]{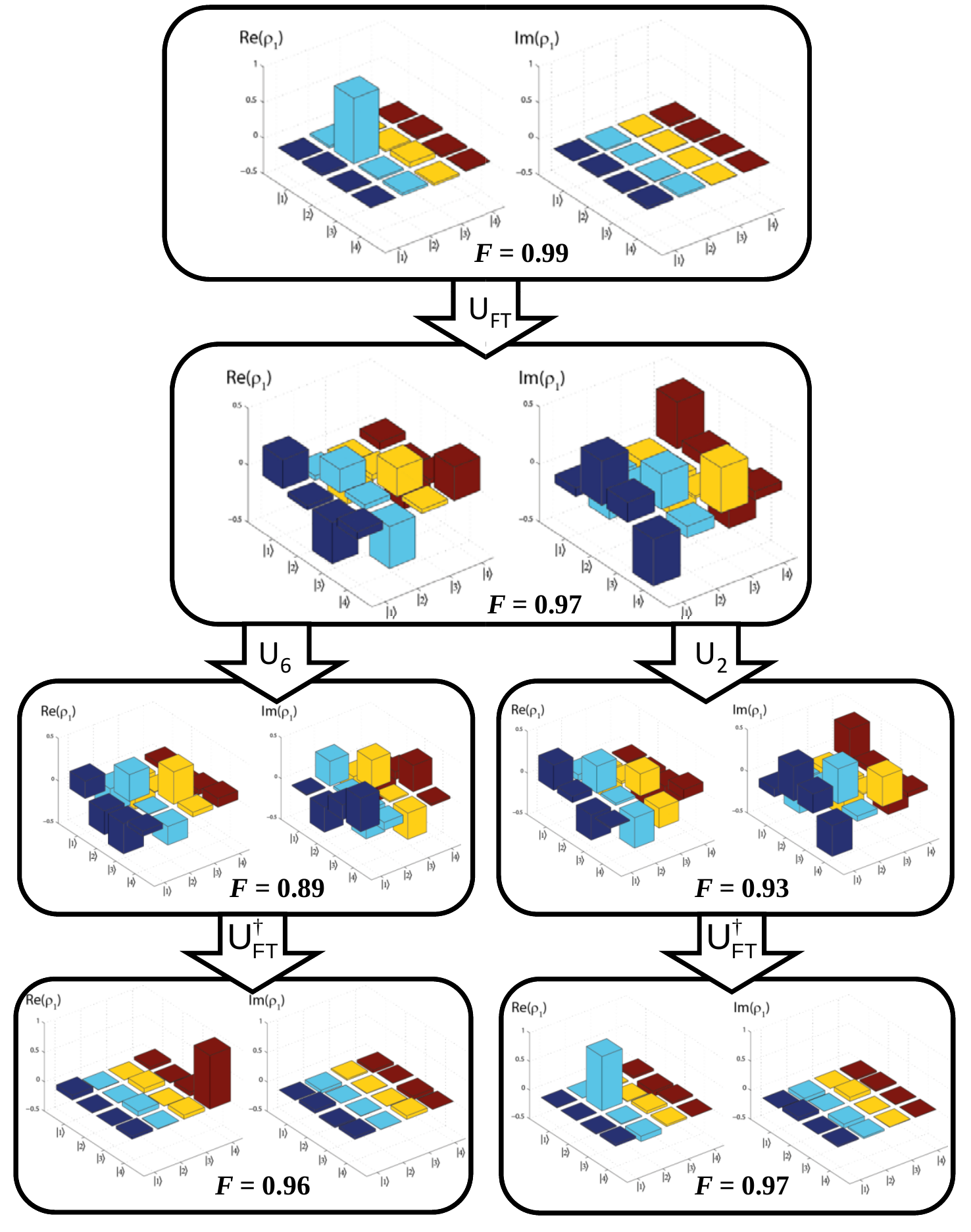} \caption{\label{Protocol} Experimental demonstration of each step of the implemented protocol. From top to bottom is shown a bar representation of the density matrix for the experimentally created initial state $|2\rangle$ obtained by quantum state tomography; the state obtained after the application of the Fourier transformation, $U_{FT}$; the results after applying the pulses that implement $U_{2}$ (right side) or $U_{6}$ (left side); the two possible outcomes of the algorithm, namely, $|2\rangle$ for positive and $|4\rangle$ for negative cyclic permutations.}
\end{figure*}

\textit{Computational task}.-- The considered quantum algorithm provides a solution to a black-box problem making use of a single four level quantum system, i.e., a single ququart. The black-box maps four possible inputs to four possible outputs after a permutation, where the eight chosen permutation functions of four objects are grouped in two ways depending on whether the permutation is positive cyclic or negative cyclic. Whereas any classical algorithm requires at least two queries to the black-box to determine the parity of the permutation, the algorithm that we experimentally realize here performs the same task with only one query, and thus provides a two to one speed-up over any classical algorithm performing the same task \cite{gedik}.

\begin{figure*}[t]
\includegraphics[scale=0.4]{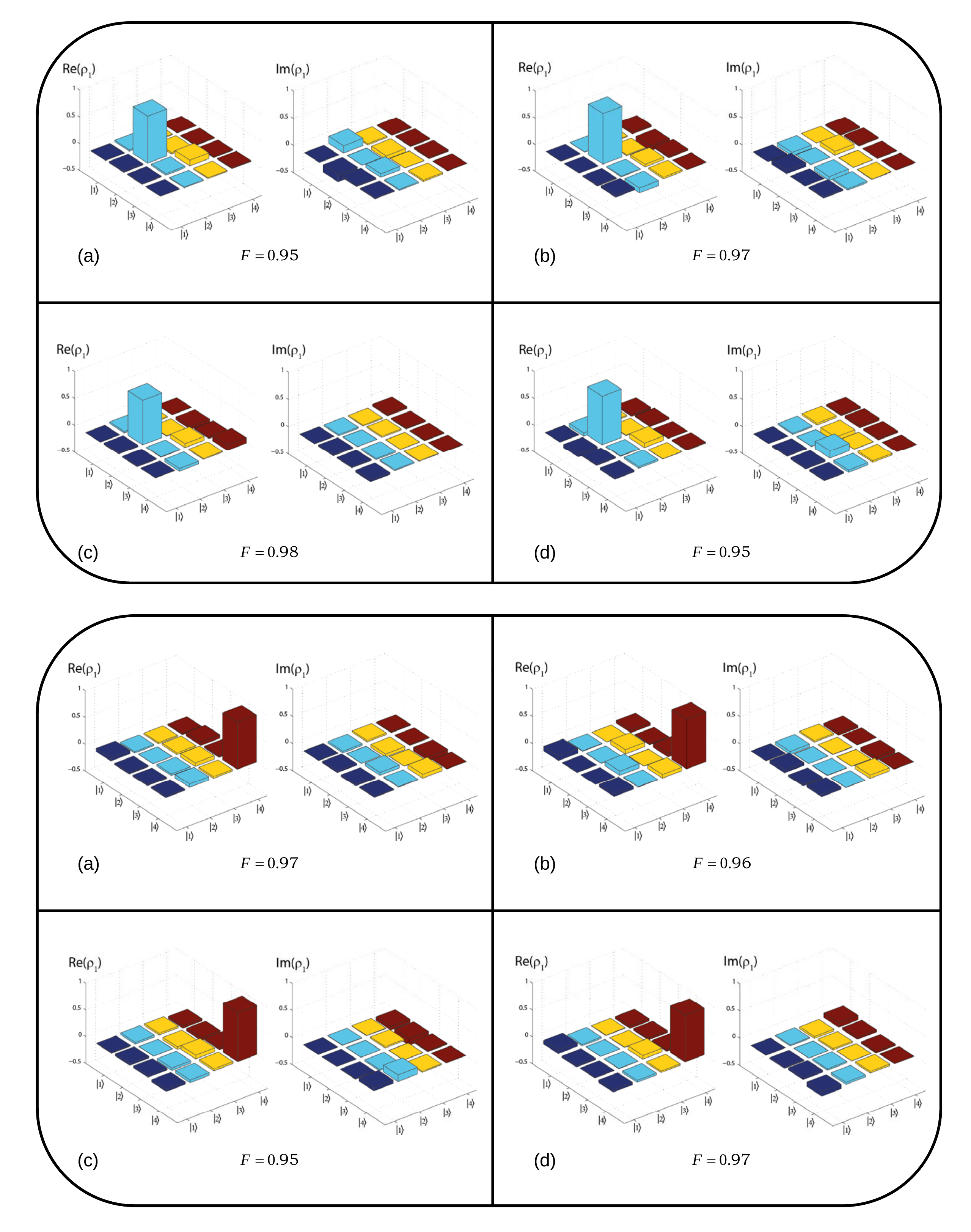} \caption{\label{Permutation} (Upper panel) Tomography results after the final step of the algorithm, together with the fidelities to the theoretical prediction, for the implementation of the positive cyclic permutation operations: (a) $U_1$, (b) $U_2$, (c) $U_3$, (d) $U_4$. (Lower panel) Tomography results after the final step of the algorithm, together with the fidelities, for the implementation of the negative cyclic permutation operations: (a) $U_5$, (b) $U_6$, (c) $U_7$, (d) $U_8$.}
\end{figure*}

We consider the eight chosen permutations of the set $\{1,2,3,4\}$, where $(1, 2, 3, 4)$, $(2, 3, 4, 1)$, $(3, 4, 1, 2)$ and $(4, 1, 2, 3)$ are positive cyclic permutations and the remaining four, namely $(4, 3, 2, 1)$, $(3, 2, 1, 4)$, $(2, 1, 4, 3)$ and $(1, 4, 3, 2)$, are negative cyclic permutations. The computational task is to determine the parity (evenness or oddness) of the permutation. Provided we treat the permutation as a function $f(x)$ defined on the set $x\in\{1,2,3,4\}$, it is required to evaluate $f(x)$ at least twice to determine the parity of the permutation classically.

Considering a quantum solution to the same task, let us define the basis vectors of the ququart as $|1\rangle=(1, 0, 0, 0)^T$, $|2\rangle=(0, 1, 0, 0)^T$, $|3\rangle=(0, 0, 1, 0)^T$, and $|4\rangle=(0, 0, 0, 1)^T$. We first create a ququart state consisting of a superposition of all four basis vectors. In order to do this, we apply the Fourier transformation
\begin{align}
 U_{FT} &= \frac{1}{2} \begin{pmatrix} 1 & 1 & 1 & 1 \\ 1 & i & -1 & -i \\ 1 & -1 & 1 & -1 \\ 1 & -i & -1 & i \end{pmatrix},
\end{align}
to the state $|2\rangle$ and obtain
\begin{align}
|\psi_2\rangle=(|1\rangle+i|2\rangle-|3\rangle-i|4\rangle)/2.
\end{align}

Starting from $(1, 2, 3, 4)$, the unitary matrices that correspond to the positive cyclic permutations $(1, 2, 3, 4)$, $(2, 3, 4, 1)$, $(3, 4, 1, 2)$, and $(4, 1, 2, 3)$ are given as
\begin{align}
 U_1 &= \begin{pmatrix} 1 & 0 & 0 & 0 \\ 0 & 1 & 0 & 0 \\ 0 & 0 & 1 & 0 \\ 0 & 0 & 0 & 1 \end{pmatrix}, &
 U_2 &= \begin{pmatrix} 0 & 0 & 0 & 1 \\ 1 & 0 & 0 & 0 \\ 0 & 1 & 0 & 0 \\ 0 & 0 & 1 & 0 \end{pmatrix}, & \\ \nonumber
 U_3 &= \begin{pmatrix} 0 & 0 & 1 & 0 \\ 0 & 0 & 0 & 1 \\ 1 & 0 & 0 & 0 \\ 0 & 1 & 0 & 0 \end{pmatrix}, &
 U_4 &= \begin{pmatrix} 0 & 1 & 0 & 0 \\ 0 & 0 & 1 & 0 \\ 0 & 0 & 0 & 1 \\ 1 & 0 & 0 & 0 \end{pmatrix}, &
\end{align}
respectively, and they map $|\psi_2\rangle$ to $|\psi_2\rangle$, $-i|\psi_2\rangle$, $-|\psi_2\rangle$, and $i|\psi_2\rangle$. On the other hand, the unitary matrices that perform the negative cyclic permutations $(4, 3, 2, 1)$, $(3, 2, 1, 4)$, $(2, 1, 4, 3)$ and $(1, 4, 3, 2)$ are given as
\begin{align}
 U_5 &= \begin{pmatrix} 0 & 0 & 0 & 1 \\ 0 & 0 & 1 & 0 \\ 0 & 1 & 0 & 0 \\ 1 & 0 & 0 & 0 \end{pmatrix}, &
 U_6 &= \begin{pmatrix} 0 & 0 & 1 & 0 \\ 0 & 1 & 0 & 0 \\ 1 & 0 & 0 & 0 \\ 0 & 0 & 0 & 1 \end{pmatrix}, & \\ \nonumber
 U_7 &= \begin{pmatrix} 0 & 1 & 0 & 0 \\ 1 & 0 & 0 & 0 \\ 0 & 0 & 0 & 1 \\ 0 & 0 & 1 & 0 \end{pmatrix}, &
 U_8 &= \begin{pmatrix} 1 & 0 & 0 & 0 \\ 0 & 0 & 0 & 1 \\ 0 & 0 & 1 & 0 \\ 0 & 1 & 0 & 0 \end{pmatrix}. &
\end{align}
respectively, and they map $|\psi_2\rangle$ to $-i|\psi_4\rangle$, $-|\psi_4\rangle$, $i|\psi_4\rangle$, and $|\psi_4\rangle$, where $|\psi_4\rangle=U_{FT}|4\rangle$ . After each one of these unitary transformations, the algorithm is concluded with the application of the inverse Fourier transform, $U_{FT}^{\dagger}$ to obtain $|2\rangle$ for positive and $|4\rangle$ for negative cyclic permutations. Thus, the algorithm determines the parity of the permutation with only a single evaluation.

\textit{Experimental results}.-- The density matrix of a room temperature NMR system can be written as  $\rho=\frac{1}{4}\mathbb{I}_{4}+\epsilon\Delta\rho$, where $\epsilon=\hbar\omega_L /4k_BT \sim 10^{-5}$ is the ratio between the magnetic and thermal energies, $\omega_L$ is the Larmor frequency, $k_B$ is the Boltzmann constant and $T$ the temperature \cite{abra,ivan}. Measurements and unitary transformations only affect the traceless deviation matrix $\Delta\rho$, which contains all the available information about the state of the system. Unitary transformations over $\Delta\rho$ are implemented by radio frequency pulses and/or evolutions under spin interactions, with an excellent control of rotation angle and direction. The full characterization of $\Delta\rho$ can be achieved using many available quantum state tomography protocols \cite{long2001,leskowitz2004,teles2007}. Note that since in NMR experiments only the deviation matrix is detected, density matrix elements are expressed in units of $\epsilon$.

The implementation of the algorithm using a ququart is achieved using a spin--$\frac{3}{2}$ NMR quadrupolar system. A spin--$\frac{3}{2}$ in the presence of an uniaxial electric field gradient and a strong static magnetic field is a four level system, corresponding to the four possible $m_z$ values of the spin--$\frac{3}{2}$, with the Hamiltonian
\begin{align}
H = -\hbar\omega_{L} I_{z} + \frac{\hbar\omega_{Q}}{6}\,(3\,I_{z}^{2} - \bold{I}^{2}),
\end{align}
where $\omega_{L}$ is the Larmor frequency, $\omega_{Q}$ is the quadrupolar frequency ($|\omega_L| \gg |\omega_Q|$), $I_{z}$ is the $z$ component of the nuclear spin operator and $\bold{I}$ is the total nuclear spin operator \cite{abra}. The eigenstates of the system are represented by $\left|3/2\right\rangle$, $\left|1/2\right\rangle$, $\left|-1/2\right\rangle$, and $\left|-3/2\right\rangle$,  here indexed as $|1\rangle$, $|2\rangle$, $|3\rangle$, $|4\rangle$. This is experimentally realized with sodium nuclei, $^{23}$Na, in a lyotropic liquid crystal sample at room temperature. The sample is prepared with 20.9 wt\% of SDS (95\% of purity), 3.7 wt\% of decanol, and 75.4 wt\% of deuterium oxide, by following the procedure in Ref. \cite{amostra}. The $^{23}$Na NMR experiments are performed in a 9.4-T VARIAN INOVA spectrometer using a 5-mm solid-state NMR probe head at $T = 25^{º}$C. We obtain the quadrupole frequency $\nu_{Q} = \omega_{Q} / 2 \pi = 10$ kHz.

The initial state is prepared from the thermal equilibrium state using a time averaging procedure based on numerically optimized radio frequency (rf) pulses generally dubbed as the strong modulating pulses (SMP) \cite{fortunato,teles2007}. The technique consists of using blocks of concatenated rf pulses, with amplitudes, phases and durations optimized to provide a state preparation such that density matrix is given by $\rho=\frac{(1-\epsilon)}{4}\,\mathbb{I}_{4}+\epsilon\rho_1=\frac{(1-\epsilon)}{4}\,\mathbb{I}_{4}+\epsilon|i\rangle\langle i|$, where $\rho_1$ is a trace one density matrix corresponding to the state $|i\rangle\langle i|$ defined by the optimized SMP pulses \cite{ivan}. The quantum gates, which are part of the quantum protocol, are also realized using these SMP optimized pulses. Since NMR measurements are not sensitive to the identity part of the density matrix, the term $|i\rangle\langle i|$ is manipulated and read-out selectively.

In Fig. \ref{Protocol}, we show a bar representation of the density matrices obtained by quantum state tomography after the steps of the protocol. These steps are implemented as follows: ($i$) we implement the SMP optimized gate $U_{FT}$ to the initial state $|2\rangle$ to obtain $|\psi_2\rangle$; ($ii$) we implement the SMP optimized gate $U_iU_{FT}$ for $i = 2, 6$ to the initial state again; ($iii$) finally, starting once more from the initial state, we implement the SMP optimized gate $U^{\dagger}_{FT}U_iU_{FT} $ for $i = 2, 6$ to obtain either $|2\rangle$ or $|4\rangle$ as an outcome of the algorithm. The fidelities to the theoretical prediction is shown at each step of Fig. \ref{Protocol}. Moreover, Fig. \ref{Permutation} experimentally confirms that the algorithm works as intended for all eight possible permutations.

\textit{Discussion}.-- In summary, using a quadrupolar NMR setup, we have experimentally demonstrated that the proposed algorithm works as claimed, that is, it deterministically decides whether a given permutation, from a set of eight possible functions, of four objects is positive or negative cyclic with a single query to the black-box. Noting that the same computational task requires at least two queries for classical algorithms, it is clear that quantum mechanics provides a two to one speed-up here \cite{gedik}.

It might be argued that the considered computational task is not of great importance. However, it shows the power of quantum computation in a strikingly simple way. In fact, this is one of the simplest known quantum algorithms. Despite its simplicity, the origin of the speed-up remains yet unclear. It is evident that quantum correlations do not play any role in the solution of the computational task since a single quantum system is considered. Moreover, the quantum algorithm that we study brings speed-up starting from a qutrit, and is trivial for a qubit. Considering the fact that an uncorrelated but contextual system is exploited, it is possible that contextuality might be playing a role here \cite{howard}. Regardless, the true resource behind the power of this algorithm remains as an open question.

\textit{Acknowledgements}.-- Financial support was provided by Brazilian agencies CNPq, CAPES, FAPESP, INCT-Informa\c{c}\~ao Qu\^antica, and the Turkish agency TUBITAK under Grant No. 111T232.

\textit{Note added}.-- After the completion of this work, we became aware of a subsequent work \cite{dogra}, also performing the implementation of the same quantum algorithm but in a different NMR system.

\end{document}